\documentclass[11pt]{article}

\usepackage{fullpage}

\usepackage{latexsym}
\usepackage{amssymb}
\usepackage{amsmath}
\usepackage{enumerate}
\usepackage{mathtools}
\usepackage{verbatim}
\usepackage{color}

\usepackage{hyperref}

\def\e{\mbox{$\epsilon$}}

\def\X{\mbox{$\cal X$}}

\def\B{\mbox{$\cal B$}}

\def\L{\mbox{$\cal L$}}

\def\F{\mathbb F}
\def\CN{\mathbb C}

\def\cmp{\mbox{$\mathbb C$}}

\def\e{\mbox{$\epsilon$}}

\def\F{\mathbb F}
\def\Z{\mathbb Z}

\def\Z{\mathbb Z}

\def\B{\mbox{$\cal H$}}
\def\v{\mbox{$\upsilon$}}

\def\finito{{\hspace*{\fill}  \mbox{$\blacksquare $}}}

\def\defeq{\stackrel{\mathrm{def}}{=}}

\def\ceil#1{\left\lceil #1 \right\rceil}
\def\floor#1{\left\lfloor #1 \right\rfloor}

\newtheorem{theorem}{Theorem}[section]
\newtheorem{lemma}[theorem]{Lemma}
\newtheorem{definition}[theorem]{Definition}
\newtheorem{corollary}[theorem]{Corollary}

\def\X{\chi}

\def\e{{{\epsilon}}}

\def\ostar{\circledast}
\def\EX{\mathbb E}

\begin{document}
\author{Louay Bazzi
  \footnote{
  Department of Electrical and Computer Engineering, American University of Beirut, 
  Beirut, Lebanon. E-mail: louay.bazzi@aub.edu.lb.}
  \footnote{Research supported by MSFEA URB grant, American University of Beirut.
    }
}
\title{On the covering radius of small codes versus dual distance}

\maketitle
\abstract{
  Tiet\"{a}v\"{a}inen's upper and lower bounds  assert that for  block-length-$n$ linear codes with dual distance $d$, the covering radius $R$ is at most  $\frac{n}{2}-(\frac{1}{2}-o(1))\sqrt{dn}$ and  typically  at least $\frac{n}{2}-\Theta(\sqrt{dn\log{\frac{n}{d}}})$.
  The   gap  between  those bounds  on $R -\frac{n}{2}$ is an $\Theta(\sqrt{\log{\frac{n}{d}}})$ factor  related to the      gap between the worst covering radius given $d$ and  the sphere-covering bound.
  Our focus in this paper is 
   on the case when $d = o(n)$, i.e.,  when  the  code size is subexponential and the gap is    $w(1)$. 
  We show that up to a  constant,    the   gap can be eliminated by relaxing   the covering  requirement to allow for missing   $o(1)$ fraction of points.   Namely, 
  if the  dual distance $d = o(n)$, then   for   sufficiently large $d$, 
  almost all  points can be covered with  radius 
 $R\leq\frac{n}{2}-\Theta(\sqrt{dn\log{\frac{n}{d}}})$.
  Compared to random linear codes, our bound   on $R-\frac{n}{2}$ is  asymptotically tight up to a factor less than $3$. We give applications to dual BCH codes. 
  The proof  builds on the author's previous work   on the weight distribution of cosets of linear codes,  which we  simplify in this paper and extend 
  from codes to probability distributions on $\{0,1\}^n$, thus enabling  the  extension of  the above result  to $(d-1)$-wise independent distributions.  
}


\section{Introduction}\label{intros}

  The {\em covering radius} of  a subset $C$ of the Hamming cube $\{0,1\}^n$  is the minimum $r$  such that
  any vector in $\{0,1\}^n$ is within Hamming distance at most $r$ from
  $C$.  Throughout the paper, $n\geq 1$ is an integer and 
  the  {\em Hamming weight} of a vector $x\in \{0,1\}^n$, which we  denote by $|x|$,  is the  number of nonzero  coordinates of $x$.
  If $r\geq 0$ is a real number
  and $x\in \{0,1\}^n$, let $\B_n(x;r)$ be 
  the {\em Hamming ball} of radius $r$ centered at $x$, i.e.,  
  $\B_n(x; r) = \{x\in \{0,1\}^n~:~ |x+y|\leq r\}$, where $+$ is addition modulo $2$.  
  If $C$ is a subset of $\{0,1\}^n$, let $\B_n(C; r)$ be the 
  the $r$-neighborhood of $C$, i.e., $\B_n(C; r) = \cup_{x\in C} \B_n(x; r)$. 
  Thus, the covering radius of  $C$ is the minimum $r$ such that 
  $\B_n(C; r)  =\{0,1\}^n$.   See \cite{CC97} for a comprehensive reference on covering codes.

    In 1990, Tiet\"{a}v\"{a}inen derived an upper bound on the
    covering radius $R$ of a block-length-$n$ linear code $C$ 
    in terms of only its  minimum dual distance $d$.
        The {\em minimum distance} of a non-zero $\F_2$-linear\footnote{$\F_2$ is
        the finite field  structure on $\{0,1\}$.} code 
  is the minimum Hamming weight of a nonzero codeword.

    \begin{theorem}[Tiet\"{a}v\"{a}inen        \cite{T90,T91}]\label{trth90}
      {\bf (Upper bound on covering radius of codes in terms of  dual distance)} 
  Let  $C\subset \F_2^n$ be an $\F_2$-linear code whose dual has minimum distance  $d\geq 2$.
  Then   the covering radius $R$ of $C$ is at most
  \[
  \left\{\begin{array}{ll}
    \frac{n}{2} - \sqrt{s(n-s)} + s^{1/6} \sqrt{n-s} & \mbox{if $d=2s$ is even} \\
    \frac{n}{2} - \sqrt{s(n-1-s)} + s^{1/6} \sqrt{n-1-s} - \frac{1}{2} & \mbox{if $d=2s+1$ is odd.} 
  \end{array}\right.
  \]
\end{theorem}
  Tiet\"{a}v\"{a}inen's argument is based on studying the dual linear program in the context 
  of Delsarte's linear programming framework \cite{Del73a}.
  In particular, Tiet\"{a}v\"{a}inen 
  proved   Theorem \ref{trth90} 
  by establishing  the existence  of certain univariate 
  low degree polynomials
  constructed from  Krawtchouk polynomials.

Prior to Tiet\"{a}v\"{a}inen's work,  the relation between the covering radius    and dual distance was investigated
in  \cite{Del73a}- \cite{Sole90}.   
In the $d = \Theta(n)$ regime, Tiet\"{a}v\"{a}inen's bound   was later improved 
in a sequence of works      
\cite{SS93} - \cite{AB02} 
(see also \cite{FL95}).     
 For  small values of $d$ including the $d = o(n)$ regime, 
 it is still the best known upper bound.  
  In 
  \cite{Baz17}, 
we showed that  for $d\leq\frac{n^{1/3}}{\log^2{n}}$, 
Tiet\"{a}v\"{a}inen's bound on $R-\frac{n}{2}$ is asymptotically  
tight up to a factor of $2$ for $(d-1)$-wise 
independent probability distributions on $\{0,1\}^n$,  of which 
linear codes with dual distance $d$ are special cases.

 By combining the sphere-covering bound and  Gilbert-Varshamov bound, Tiet\"{a}v\"{a}inen \cite{T91}
 established also a  simple lower bound on the covering radius as function of the dual distance. 
 For comparison purposes, 
 we need the following  version of the lower bound tailored to the small $d$ regime. 
   \begin{lemma}\label{GVCSC}
     {\bf (Small codes version of Tiet\"{a}v\"{a}inen's lower bound on the covering radius in terms of  dual distance)}
           Let $n\geq 1$ be an integer and 
           $ n \leq K \leq 2^{n-1}$ be  an integer power of $2$.  
              Then, almost all 
                    $\F_2$-linear codes $C\subset \F_2^n$ of size $K$ have 
          covering radius 
$R \geq  
\frac{n}{2} - \Theta(\sqrt{  dn  \log{\frac{n}{d}}})$, 
where $d$ is the minimum distance  of $C^\bot$. 
More specifically, 
all but at most   $\frac{1}{n}$ fraction of 
         $\F_2$-linear codes $C\subset \F_2^n$ of size $K$ have 
          covering radius 
\[
R \geq  
\frac{n}{2} - \sqrt{
  \frac{dn}{2}  \log{\frac{en}{d}} +   n \log{(n+1)}
}.        \]
   \end{lemma}
   For a proof of Lemma \ref{GVCSC}, see Lemma \ref{GVCSCA} with $\varepsilon=0$. Note that throughout the paper $\log$ means $\log_e$.

     The difference between  Tiet\"{a}v\"{a}inen's upper and lower bounds on $R -\frac{n}{2}$ is 
     a $\Theta(\sqrt{\log{\frac{n}{d}}})$ factor.      
   The focus of this brief paper is on linear codes with dual distance $d = o(n)$, which corresponds to 
  the case when the code size is   subexponential and the factor $\Theta(\sqrt{\log{\frac{n}{d}}})$ grows with $n$.
  Our study  is motivated by this gap 
which  is  related to 
the  gap between
the typical   and the worst possible 
covering radius given  $d$.
In what follows, we highlight the gap by comparing the covering radius of dual BCH codes with the typical covering radius of  linear codes of the same size.

It  follows from the work of Cohen  and Blinovskii that the typical covering radius of linear codes  achieves the sphere-covering bound.
Cohen showed that there are linear codes up to  the sphere-covering bound:

\begin{theorem} \label{cohthm}  {\bf (Cohen \cite{coh83}; see also \cite[Chapter~12]{CC97})}~{\bf (Linear codes up to the sphere-covering bound)}  
    For any $n\geq 1$   and $0< R \leq n$,  
    there exists an $\F_2$-linear code $C\subset \F_2^n$ of covering radius   $R$ and dimension 
\[
k \leq \ceil{ \log_2{  \frac{n (\log{2})}{\v_n( R)}}},
\] 
where $\v_n( R)$ is the probability with respect to the uniform distribution of the radius-$R$ Hamming ball $\B_n(0;R)$.  
\end{theorem}
  Later, Blinovskii
  \cite{Bli87,Bli90}
  showed that almost all linear codes achieve  the sphere-covering bound.
  See also   
  \cite[Chapter~12]{CC97} and the references therein.

    To  illustrate the gap in the case of dual BCH code,  
    we  need the following immediate corollary to  Cohen's theorem customized to  small codes. 
  We include a   proof in Appendix \ref{appb} for completeness.
\begin{corollary} {\bf (Explicit     version for small codes)}\label{cohcor} 
    If $n\geq 1$,  $s> 1$, and  $s = o(\frac{n}{\log{n}})$, then 
     for $n$ large enough, 
    there exists an $\F_2$-linear code $C\subset \F_2^n$ of dimension at most     $\ceil{s \log_2{n}}$  and 
    covering radius 
      $R \leq \frac{n}{2} - \sqrt{ \frac{(s-1) n\log{n}}{2+o(1)}}$.  

    More specifically,
  for each $\e>0$, there exists $\delta>0$ such that the following holds. 
    Let $n\geq 1$ be an integer and $s> 1$ be  such that $s\log_2{n} \leq \delta n$. 
    Then for $n$ large enough, 
    there exists an $\F_2$-linear code $C\subset \F_2^n$ of dimension at most     $\ceil{s \log_2{n}}$  and 
    covering radius 
\[
R \leq \frac{n}{2} - \sqrt{\frac{ (s-1) n\log{n} }  {2+\e}} + \sqrt{2n}+2. 
\]
\end{corollary}
Before moving to the next section, we note that a  related   work is
    an explicit construction  due to Alon -- 
  attributed  to  Alon by Rabani and Shpilka \cite{RS10} -- of  polynomial size codes of covering radius $\frac{n}{2}-c\sqrt{n\log n}$, for any constant $c$.

\subsection{The gap in the case of dual BCH codes}\label{probdef}
  Consider the block-length-$n$ dual BCH code $C = BCH(s,m)^\bot$, where 
  $m\geq 2$ and $s\geq 1$ are   integers such that
  $2s-2< 2^{m/2}$, i.e., $s < \frac{1}{2}\sqrt{n+1}+1$, and $n = 2^m -1$. 
  The dimension of $C$ is $k = sm = s \log_2{(n+1)}> s \log_2{n} $, the 
  minimum distance of $C^\bot$  is at least $d=2s+1$,    and the    covering radius $R$ of $C$ satisfies:  
  \begin{eqnarray}
      R &\leq& \frac{n}{2} - (1-o(1))\sqrt{sn}
      \label{BCHde11}\\
    R &\geq&         \frac{n}{2} - (s-1)\sqrt{n+1}-\frac{1}{2}. \label{BCHde12}
  \end{eqnarray} 
  The upper bound (\ref{BCHde11}) is   Tiet\"{a}v\"{a}inen's bound (Theorem \ref{trth90})  and 
  the lower bound  (\ref{BCHde12}) 
     is  Weil-Carlitz-Uchiyama bound (see Section \ref{dualBCH}). 
     Applying Corollary \ref{cohcor} to linear codes of 
     dimension comparable to the dimension $k$ of the dual BCH code $C$,  
     we get that 
     there exists an $\F_2$-linear code $C'\subset \F_2^n$ of dimension $k' \leq \ceil{s \log_2{n}}\leq k$  and 
     covering radius    
  \begin{equation}\label{BCHde2}
    R' \leq \frac{n}{2} - \sqrt{\frac{ (s-1) n\log{n} }  {2+o(1)}}. 
  \end{equation} 
  Comparing  (\ref{BCHde11}) and (\ref{BCHde2}), we see that the upper bound on $R-\frac{n}{2}$ in
  (\ref{BCHde11})
  is worse than that in (\ref{BCHde2}) by a factor of $\Theta(\sqrt{\log{n})}$.
  The same factor appears  if we compare the lower bound (\ref{BCHde12})
  with the upper bound   (\ref{BCHde2}) when $s = \Theta(1)$. 
  That is, in the $s = \Theta(1)$ regime, while
   the actual covering radius
   of $BCH(s,m)^\bot$ is $R =\frac{n}{2} -  \Theta(\sqrt{n})$,
   linear codes of smaller dimension have covering radius
   $R =\frac{n}{2} -  \Theta(\sqrt{n\log{n}})$.

  \subsection{Summary of results}
  For dual BCH codes $BCH(s,m)^\bot$, where $s\geq 3$ and  $2s-2< 2^{m/2}$, 
  we show that 
  the  $\Theta(\sqrt{\log{n}})$ gap can be eliminated
  by relaxing   the covering requirement: 
  instead of covering all the vectors in $\{0,1\}^n$, 
  we can guarantee covering all but   $o(1)$ fraction of them  
  with radius   $\frac{n}{2} -  \Theta(\sqrt{sn\log{n}})$.  
    More generally, we show that if a linear code has  dual minimum distance at least $d$,  where
    $d = o(n)$, 
  then for   sufficiently large $d$, 
  almost all  points can be covered with  radius     $R\leq\frac{n}{2}-\Theta(\sqrt{dn\log{\frac{n}{d}}})$.   
  This bound   on $R-\frac{n}{2}$  asymptotically  matches    Tiet\"{a}v\"{a}inen lower bound up to factor less than $3$.
  It also asymptotically  matches up to the same factor an adaptation  of Tiet\"{a}v\"{a}inen lower bound to  almost-all-coverings, i.e., 
  compared to random linear codes, it is tight   up to a constant factor less than $3$.  
  The proof  builds on the author's previous work  
  on the weight distribution of cosets of linear codes with given 
  bilateral minimum distance \cite{Baz15}.   
    The {\em bilateral minimum distance} of a non-zero $\F_2$-linear code $D$ is the maximum $b$ such that  all nonzero codewords have weights between $b$ and $n-b$, i.e., 
$b\leq |z|\leq n-b$, for each nonzero  $z\in D$.

      We also simplify in this paper a part of the proof in \cite{Baz15}  which makes it possible to extend the results in \cite{Baz15} as well as
      the above results  from codes to probability distributions. 
      In particular, we  extend the above results  on the almost-all covering radius from
      codes with dual distance $d$ to $(d-1)$-wise independent distributions, of which linear codes with dual distance $d$ are special cases. 
        A probability distribution $\mu$ on $\{0,1\}^n$ is called {\em $k$-wise independent} if for       $(x_1,\ldots, x_n)\sim \mu$, each $x_i$
 is equally likely to be $0$ or $1$ and any $k$ of the $x_i$'s are statistically independent \cite{Lub85,Vaz86}.  
  Linear codes with dual distance at least $k+1$ are special cases of $k$-wise independent distributions.  Namely, if $\mu$ is 
  uniformly distributed on an $\F_2$-linear code $C\subset \F_2^n$,    then $\mu$ being $k$-wise independent is equivalent to $C$ having
   dual minimum distance at least $k+1$. 
      Note that the {\em covering radius of a probability distribution}    
      on $\{0,1\}^n$ is interpreted  as
      the covering radius of  its support.

We elaborate below on the  results in the case of linear codes. Their extensions to distributions are presented in Section \ref{extenap}. 
\begin{definition} {\bf (Almost-all covering)} Let $0\leq \varepsilon \leq 1$.
  The {\em $\varepsilon$-covering radius} of  a subset $C$ of the Hamming cube $\{0,1\}^n$  is the minimum $r$
  such that the fraction of points of the Hamming cube not contained in the 
  $r$-Hamming-neighborhood $\B_n(C; r)$ of $C$ is at most $\varepsilon$.
\end{definition}
Thus the covering  radius corresponds to $\varepsilon=0$.  The notion of  almost-all-covering goes back to  the argument of  Blinovskii  \cite{Bli87,Bli90} 
 in his  proof    that almost all linear codes  achieve the sphere-covering bound.

  First we establish the following nonasymptotic bound. 
  \begin{theorem}\label{brute}
    {\bf (Upper bound on the almost-all-covering radius of small codes in terms of dual distance)}
        Let $ C\subsetneq \F_2^n$  be an $\F_2$-linear code whose dual $C^\bot$ has minimum distance at least $d$,
where $d\geq 7$  be  an odd integer.  
  If  $R>  0$, 
  then 
  the fraction of points in Hamming cube not covered by $\B_n(C; R)$   is at most
\[
\varepsilon= \frac{d}{ \v_{n+d}(R)     } \left(e\log{\frac{n+d}{d-1}}\right)^{\frac{d-1}{2}}\left(\frac{d-1}{n+d}\right)^{ \frac{d-5}{4}},  
\]
where $\v_{n+d}( R)$ is the probability  with respect to the uniform distribution of the radius-$R$ Hamming ball  $\B_{n+d}(0;R)$ in $\{0,1\}^{n+d}$.  That is,
if $\varepsilon \leq 1$, then the  $\varepsilon$-covering radius of $C$ is at most $R$.
  \end{theorem}
  The proof of Theorem \ref{brute}   builds on  \cite{Baz15}, where it is shown that for
            an $\F_2$-linear code $Q$ with  dual bilateral minimum distance at least $b$,  the average $L_1$-distance  between the 
            weight distribution of a random cosets of $Q$ and the binomial distribution  decays quickly in $b$, and namely, it is bounded by
            $b\left(e\log{\frac{n}{b-1}}\right)^{\frac{b-1}{2}} \left(\frac{b-1}{n}\right)^{\frac{b-5}{4}}$, if $b\geq 7$ is odd.  
            The proof of Theorem \ref{brute} boils down to using Markov Inequality and applying the above result 
            to the block-length $n+d$ code $Q$ constructed from $C$  by appending  $d$ independent bits to $C$. This simple construction  
            turns the 
            lower bound $d$ on the minimum distance of $C^\bot$ into a lower bound on the
            bilateral minimum distance  of $Q^\bot$.
  
            Then, based on the entropy estimate of the binomial coefficients,  we conclude the following bound    in the $d = o(n)$ regime.  
 \begin{corollary}\label{maincor1}
    {\bf (Explicit asymptotic version)}    
                            Let $C\subsetneq \F_2^n$  be an $\F_2$-linear code whose dual $C^\bot$ has minimum distance at least $d$.

                            If $d = o(n)$,   then for   sufficiently large $d$, 
  the $o(1)$-covering radius of $C$ is at most 
  $\frac{n}{2}-\Theta(\sqrt{dn\log{\frac{n}{d}}})$.

  More specifically, if $d\geq 7$ is an odd integer such that $d =o(n)$, 
                     then, for sufficiently large $n$, the $\left(\frac{d-1}{n}\right)^{\frac{d-5}{13}}$-covering radius of $C$ is at most $R = \frac{n}{2}-\Delta$, where
                     $$\Delta  = \sqrt{ \frac{1}{13}(d-5)n\log{\frac{n}{d-1}}}.$$
  \end{corollary}
  Comparing the bounds on $R-\frac{n}{2}$ in Corollary \ref{maincor1} and Lemma \ref{GVCSC}, we see that the 
  guarantee given by Corollary \ref{maincor1} on $R - \frac{n}{2}$ 
  is asymptotically
  not far from Tiet\"{a}v\"{a}inen's lower bound on the covering radius of random linear codes by more than a factor of 
  $\sqrt{\frac{13}{2}} \approx 2.55 < 3$.  Actually, for almost-all-coverings, the upper bound of 
  Corollary \ref{maincor1} is asymptotically tight up to the same factor in comparison to random linear codes. 
  This follows from the following simple variation of
  Lemma \ref{GVCSC}:  
     \begin{lemma}\label{GVCSCA}
       {\bf (Variation of Tiet\"{a}v\"{a}inen's lower bound: lower bound on the  almost-all-covering                   radius of small codes in terms of  dual distance)}
       Consider any  $0\leq \varepsilon < 1$ and let             $n\geq 1$  be an integer   and            $ n \leq K \leq 2^{n-1}$  be  an integer power of $2$.
             Then, almost all   $\F_2$-linear codes $C\subset \F_2^n$ of size $K$ have  $\varepsilon$-covering radius $R \geq  \frac{n}{2} - \Theta(\sqrt{dn\log{\frac{n}{d}}  +  n \log{\frac{n}{1-\varepsilon}}})$, 
        where $d$ is the minimum distance  of $C^\bot$.          
              More specifically,
               all but at most   $\frac{1}{n}$ fraction  
        of  $\F_2$-linear codes $C\subset \F_2^n$ of size $K$ have  $\varepsilon$-covering radius         
\[
R \geq  
\frac{n}{2} - \sqrt{
  \frac{dn}{2}  \log{\frac{en}{d}} +   n \log{\frac{n+1}{1-\varepsilon}}
}.        \]
     \end{lemma}
See Appendix \ref{appa} for a proof of Lemma \ref{GVCSCA}.  
     
Applying Corollary \ref{maincor1} to the dual BCH codes $BCH(s,m)^\bot$
with $d = 2s+1$, where $s \geq 3$ so that $d \geq 7$, 
we get the following:               
               \begin{corollary} \label{dbchapp}    {\bf (Application to dual BCH codes)}      
                 Let $m\geq 2$ be an integer and $n = 2^m -1$. Let 
                 $s\geq 3$ be an integer such that $2s-2 < 2^{m/2}$, i.e.,
                 $s < \frac{1}{2}\sqrt{n+1}+1$
                 and consider the dual BCH code $C = BCH(s,m)^\bot$.  Then, 
the $o(1)$-covering radius of $C$ is at most
$\frac{n}{2}-\Theta(\sqrt{ sn\log{\frac{n}{s}}}).$                
More specifically,  for sufficiently large $n$,
the $\left(\frac{2s}{n}\right)^{\frac{2s-5}{13}}$-covering radius of $C$ is at most
$$R = \frac{n}{2}-
\sqrt{ \frac{1}{13}(2s-4)n\log{\frac{n}{2s}}}.$$
               \end{corollary}
               For instance, for $s = 3$, we have
               $R = \frac{n}{2} - \sqrt{\frac{2}{13}n\log{\frac{n}{6}}}$.  
                 Thus, for $BCH(3,m)^\bot$,  even though we need
      an $\frac{n}{2} - \Theta(\sqrt{n})$ radius to cover all  points in   $\{0,1\}^n$, we can
      cover almost all of them using an $\frac{n}{2}- \sqrt{\frac{2}{13}n\log{\frac{n}{6}}}$  
      radius.       

                       Using Cohen's
                 iterative argument  for showing the existence of linear
                 coverings up the sphere-covering bound \cite{coh83}, we conclude from Corollary \ref{maincor1} that there exists 
           a small $\ceil{\log_2{n}}$-dimensional linear code which can be added to $C$ to turn the 
           almost cover  into a total cover.
           \begin{corollary}\label{maincor2}     {\bf (Adding a small code)}  
              Let $C\subsetneq \F_2^n$  be an $\F_2$-linear code whose dual $C^\bot$ has minimum distance at least $d$, where $d\geq 7$ is an odd integer such that $d   = o(n)$.
Then there exists an $\F_2$-linear code $D$ of dimension at most $\ceil{\log_2{n}}$ such that, 
for sufficiently large $n$, the covering radius of $C+D$ is at most  $\frac{n}{2}- \Theta(\sqrt{dn\log{\frac{n}{d}}})$. 
  \end{corollary}
           See Section \ref{openprobs} for a related open problem on dual BCH codes.

           Before outlining  the rest of the paper in the next section, we highlight   additional links   with the existing literature.

  Turning an almost-all linear covering into a total covering goes back to the work of
  Blinovskii \cite{Bli87,Bli90}.

  The notion of bilateral minimum distance $b$ of a linear code is  equivalent to its  {\em width} 
$\sigma$ which is given by  $\sigma = n - 2b$. 
            For small values of $b$, it is more convenient to work with $b$ rather than $\sigma$.
            In the high rate regime, the relation between the covering radius and the dual width was  studied  by Sole  and Stokes \cite{SS93}.

                     Finally, we compare with the related work of   Navon and  Samorodnitsy     \cite{NS09},  
              who recovered the first linear programming bound 
              using a covering argument and
Fourier analysis techniques. The related result in \cite{NS09} is 
a bound  that relates the   dual distance to  the minimum radius which guarantees  covering a significant fraction of the Hamming cube. 
Namely, in terms of $\varepsilon$-covering, Corollary 1.5 in \cite{NS09} asserts that 
if $C$ is a block-length-$n$ $\F_2$-linear code with dual distance $d$, then the $(1 -\frac{1}{n})$-covering radius $R$ of $C$ is
            at  most             $\frac{n}{2}-\sqrt{d(n-d)} + o(n)$. 
            Thus, in the context of
            $(1 -\frac{1}{n})$-coverings, 
             Navon-Samorodnitsky's upper bound on $R-\frac{n}{2}$ 
is stronger than  Tiet\"{a}v\"{a}inen's upper bound (Theorem \ref{trth90})  by factor of $\sqrt{2}$.
It is however weaker than our bound in Corollary \ref{maincor1} 
by factor of  $\Theta(\sqrt{\log{\frac{n}{d}}})$ in the $d = o(n)$ regime. Also, Corollary \ref{maincor1} allows for smaller values of $\varepsilon$.

\subsection{Paper outline}

After introducing some  preliminaries  in Section \ref{termo}, we prove Theorem \ref{brute} in
Section \ref{brutep}.  In Section \ref{corprf}, we prove  Corollaries \ref{maincor1} and \ref{maincor2}.
In Section  \ref{extenap}, we extend the  results from  codes to distributions.

\section{Preliminaries}\label{termo}
In this section, we compile some notations and definitions used   throughout the paper. 
\subsection{Notations}\label{termino}

We will use the following notations as in \cite{Baz17}.
The set $\{0, \ldots, n\}$ is denoted by $[0:n]$.  
The binomial distribution on $[0:n]$ is denoted by $B_n$, i.e., $B_n(w)=\frac{1}{2^n}{\binom{n}{w}}$.
The uniform distribution on $\{0,1\}^n$ is denoted by $U_n$, i.e., $U_n(x) = \frac{1}{2^n}$, for all $x\in \{0,1\}^n$.

  Thus, in terms of the above notations, the 
  {\em  $\varepsilon$-covering radius} of  a subset $C\subset \{0,1\}^n$  is the minimum $r$
  such that $U_n(\B_n(C;r))\geq 1 - \varepsilon$.

  If $\mu$ is a  probability distribution,  $\EX_\mu$  denotes the expectation  with respect to $\mu$    and ``$x\sim \mu$''  denotes the process of sampling  a random vector $x$ according to $\mu$.

\paragraph{Weight distributions}
We will also use the following  notations as in   \cite{Baz15}. 
If $\mu$ is a probability distribution on $\{0,1\}^n$,  $\overline{\mu}$ denotes the corresponding {\em weight distribution} on $[0:n]$, 
i.e., for all $w\in [0:n]$, 
$\overline{\mu}(w)\defeq \mu(x \in \{0,1\}^n: |x|=w)$. 

If $A\subset \{0,1\}^n$,  $\mu_A$ denotes the probability distribution on $\{0,1\}^n$ uniformly distributed on $A$. Thus $\overline{\mu_A}(w)$ is the fraction of points in $A$ of 
weight $w$.

\subsection{Hamming Ball Volume}
Let $\v_n( R)$ denote the probability with respect to  the uniform distribution of the radius-$R$ Hamming ball, i.e.,
$$\v_n( R) = U_n(\B_n(0; R    )) = \sum_{w \leq R} B_n(w).$$
The proofs of Corollaries \ref{cohcor} and \ref{maincor1} use the lower bound on  $\v_n( R)$ in Lemma \ref{demltc} below.  
The lower bound  is based on the  following estimate  
of the binomial coefficients: if $n\geq 1$ is an integer and $0 <\lambda<1$ is such that $\lambda n$ is an integer, then 
\begin{equation*}
  \binom{n}{\lambda n} \geq
  \frac{2^{nH(\lambda)}}{\sqrt{8n \lambda (1-\lambda)}},
\end{equation*}
where $H(x) = - x \log_2{x} - (1-x)\log_2{(1-x)}$ is the binary entropy function (see, e.g., \cite[Lemma~2.4.2]{CC97}).  
\begin{lemma} \label{demltc}  
  For each $\e>0$, there exists $\delta > 0$ such that the following holds. 
  Let $R = \frac{n}{2} - \Delta$, where 
$\Delta>0$ is such that $\Delta \leq \delta n$.
  Then, for $n$ large enough,
  \[
  \v_n( R)   \geq  e^{-(2+\e)\frac{(\Delta+\sqrt{2n}+2)^2}{n}}. 
             \]
             \end{lemma}
    {\bf Proof:}
    With  $\lambda= \frac{1}{n}\floor{\frac{n}{2} - \Delta- \sqrt{2n}-1    }$, we have 
   $$\v_n( R) = \sum_{w \leq \frac{n}{2} -\Delta}  B_n(w) \geq \sqrt{2n}
    B_n(  \lambda n  ) \geq
    \sqrt{\frac{2n}{8n \lambda (1-\lambda)}} 2^{-n(1-H(\lambda))} \geq
    2^{-n(1-H(\lambda))}.
    $$
    Let $x=\frac{\Delta+ \sqrt{2n}+2}{n}$
    , hence  $\lambda \geq \frac{1}{2} - x$. 
    Since $H(\frac{1}{2} - x )
    = 1  - \frac{2x^2}{\log{2}}-O(x^4)$, let $\delta > 0$ so that
    $H(\frac{1}{2} - x )  \geq 1 - \frac{(2+\e)x^2}{\log{2}}$ for each
    $0 \leq x \leq 2 \delta$.
    Thus
    $$\v_n( R) \geq 2^{-n(1-H(\lambda))} \geq
    2^{-n(1-H(1-x))} \geq e^{-(2+\e)nx^2 } =
    e^{-(2+\e)\frac{(\Delta+\sqrt{2n}+2)^2}{n}}. $$
    The claim then holds for $n$ sufficiently large so that
    $\frac{\sqrt{2n}+2}{n}\leq \delta$.  
    \finito

\subsection{Fourier transform preliminaries}\label{frprelS}

  We  compile  in this section   
  harmonic analysis  preliminaries as in \cite{Baz15, Baz17}. 
See \cite[Section IV]{Baz15} for a more detailed treatment.  
The notions in this section are used in Sections \ref{limindep} and \ref{extenap}.

Consider the abelian  group structure  $\Z_2^n = (\Z/2\Z)^n$ on 
the hypercube $\{0,1\}^n$ and  the  $\cmp$-vector space $\L(\Z_2^n) = \{f: \Z_2^n\rightarrow \cmp\}$ 
endowed with the  inner product:   $$\langle f,g\rangle = \EX_{U_n} f \overline{g}= \frac{1}{2^n} \sum_{x} f(x) \overline{g(x)}.$$
The  {\em characters} of $\Z_2^n$ are $\{ \X_z \}_{z\in \Z_2^n}$, where    $\X_z:\Z_2^n\rightarrow \{-1,1\}$ is given by $\X_z(x) =  (-1)^{\langle x, z\rangle}$  and 
$\langle x, z\rangle = \sum_{i=1}^{n} x_iz_i$. 
They form an orthonormal  basis of $\L(\Z_2^n)$, i.e.,  $\langle \X_z,\X_{z'}\rangle = \delta_{z,z'}$,   for each $z,z'\in \{0,1\}^n$,  where $\delta$ is the Kronecker  delta function.

  The Fourier transform    of a function  $f\in \L(\Z_2^n)$ is the function 
$\widehat{f}\in \L(\Z_2^n)$  given by the coefficients of the unique 
 expansion of $f$ in terms of the characters: \[
f(x) = \sum_{z} \widehat{f}(z)\X_z(x) 
\mbox{ }\mbox{ }\mbox{ and } \mbox{ }\mbox{ }
\widehat{f}(z) = \langle f, \X_z\rangle  =\EX_{U_n} f \X_z.
\]

\subsection{Limited independence, Fourier transform, and bilateral limited independence}\label{limindep}

In this section, we highlight the limited independence property in the Fourier domain and we
 define the notion of bilateral limited independence.
The notions in this section are   used in Section \ref{extenap}.

Let $\mu$ be  probability distribution  on $\{0,1\}^n$.
 In terms of the  characters $\{\X_z\}_z$ of $\Z_2^n$, $\mu$ being  $k$-wise independent  is equivalent to  
  $\EX_\mu \X_z = 0$   for each nonzero $z\in \{0,1\}^n$  such that $|z|\leq k$.

      We define  the notion of bilateral $k$-wise independence to match the 
    dual bilateral minimum distance in the case of linear codes.
  Recall that, if $\mu$ is 
  uniformly distributed on an $\F_2$-linear code $C\subset \F_2^n$, i.e.,
  $\mu = \mu_C$,    then $\mu$ being $k$-wise independent is equivalent to $C$ having
  dual minimum distance at least $k+1$. 
  We call a probability distribution $\mu$ on $\{0,1\}^n$ {\em bilaterally $k$-wise independent}  if
  $\EX_\mu \X_z = 0$   for each nonzero $z\in \{0,1\}^n$  such that $|z|\leq k$ or $|z|\geq n - k$.
Thus, if $\mu$ is  uniformly distributed on an $\F_2$-linear code $C\subset \F_2^n$,  
then $\mu$ being bilaterally 
$k$-wise independent is equivalent to $C$ having
  {\em bilateral dual minimum distance} at least $k+1$.

    \subsection{Dual BCH codes}\label{dualBCH}
    For a general reference on dual BCH codes, see   \cite{MS77}.  We compile in this section some of their basic properties used in the introduction.
Let $m\geq 2$ be an integer and $n = 2^m -1$. Consider the 
 finite field $\F_{2^m}$ on  $2^m$ elements and let  
 $\F_{2^m}^\times$ be $\F_{2^m}$ excluding zero. 
Let $s\geq 1$ be an integer such that $2s-2< 2^{m/2}$, i.e., $s < \frac{1}{2}\sqrt{n+1}+1$. 
Consider the BCH code  $BCH(s,m) \subset\F_2^n$: 
      $$BCH(s,m) = \{ (f(a))_{a\in \F_{2^m}^\times} ~:~ f \in \F_{2^m}[x] \mbox{ such that } deg(f) < 2^m - 2s -1 \} \cap \F_2^{\F_{2^m}^\times}.$$
We have: 
\begin{itemize}
\item[a)] $dim(BCH(s,m)^\bot) = ms$
\item[b)] The minimum distance  of $BCH(s,m)$  is at least $2s+1$
\item[c)](Weil-Carlitz-Uchiyama Bound) For each non-zero codeword $x\in BCH(s,m)^\bot$,  we have $||x|-2^{m-1}|\leq (s-1)2^{m/2}$, hence 
$||x|-\frac{n+1}{2}| \leq (s-1)\sqrt{n+1}$. 
\end{itemize}
Let $R$ be the covering radius of dual BCH code $BCH(s,m)^\bot$.  
It follows from (c) that
\[ 
  R \geq \frac{n}{2} - (s-1)\sqrt{n+1}-\frac{1}{2}.
\]
This holds because  with $\vec{1}$  denoting the all-ones vector, we have for each $x\in BCH(s,m)^\bot$,  
$|\vec{1}+x| = n-|x|\geq n  - (\frac{n+1}{2}+(s-1)\sqrt{n+1}) = \frac{n}{2} - (s-1)\sqrt{n+1}-\frac{1}{2}$.

\section{Proof of Theorem \ref{brute}} \label{brutep}
The statement of Theorem  \ref{brute} is restated below for convenience. 
~\medskip \smallskip \\
{\bf Theorem \ref{brute}}~    {\bf (Upper bound on the almost-all-covering radius of small codes in terms of dual distance)}~{\em
    Let $ C\subsetneq \F_2^n$  be an $\F_2$-linear code whose dual $C^\bot$ has minimum distance at least $d$,
where $d\geq 7$  be  an odd integer. 
  If  $R>  0$, 
  then  
the fraction  of points in Hamming cube not covered by $\B_n(C; R)$   is at most
\[
\frac{d}{ \v_{n+d}(R)     } \left(e\log{\frac{n+d}{d-1}}\right)^{\frac{d-1}{2}}\left(\frac{d-1}{n+d}\right)^{ \frac{d-5}{4}}. 
\]                   
         }\medskip\smallskip \noindent
The  proof  builds on \cite{Baz15}: 
\begin{theorem}{\bf \cite[Corollary~3]{Baz15}}\label{wdl1} {\bf (Dual bilateral minimum distance    versus weight distribution of cosets of small codes; $L_1$-bound)}  
  Let $Q\subsetneq \F_2^n$  be an $\F_2$-linear code whose dual $Q^\bot$ has bilateral minimum distance at least $b$,
  where $b\geq 7$  is an odd integer.    
 Then 
$$\EX_{u\sim U_n} \| \overline{\mu_{Q+u}} - B_n \|_1  \leq b\left(e\log{\frac{n}{b-1}}\right)^{\frac{b-1}{2}}\left(\frac{b-1}{n}\right)^{ \frac{b-5}{4}}.$$
\end{theorem}
See Section \ref{termino} for  weight distribution  notations. 
  At a high level, the proof of Theorem \ref{wdl1} uses Fourier analysis techniques to establish a mean-square-error identity. Then the  argument proceeds by estimating  the dual linear program in the context 
  of Delsarte's linear programming framework \cite{Del73a}. The dual estimate is based on Taylor
  approximation of the exponential function.

Using Markov Inequality\footnote{If $X$ is a random variable taking nonnegative values and $a>0$, then  the probability that $X \geq a$ is at most $\frac{\EX[X]}{a}$.}, we obtain the following corollary to Theorem \ref{wdl1}. 
\begin{corollary}\label{corwcdr} {\bf (Upper bound on the almost-all-covering radius of small codes in terms of dual bilateral distance)} 
    Let $Q\subsetneq \F_2^n$  be an $\F_2$-linear code whose dual $Q^\bot$ has bilateral minimum distance at least $b$,
    where $b\geq 7$  is an odd integer.  
  If $R>  0$, 
  then 
  the fraction $p$ of points in the Hamming cube not covered by $\B_n(Q; R)$    is at most
      \[
      \frac{b}{ \v_n(R)     } \left(e\log{\frac{n}{b-1}}\right)^{\frac{b-1}{2}}\left(\frac{b-1}{n}\right)^{ \frac{b-5}{4}}.
  \]
\end{corollary}
    {\bf Proof:} Choose a uniformly random $u \in \{0,1\}^n$, thus  $p$ is the probability that $Q \cap \B_n(u;  R ) =\emptyset$.
    Let $f:\{0,1\}^n\rightarrow \{0,1\}$ be the indicator function of $\B_n(0;   R)$, i.e., $f(x) = 1$ if  $|x| \leq R$ and $f(x)=0$ otherwise. 
    If $Q \cap \B_n(u;  R   ) =\emptyset$,  i.e.,    $\EX_{Q+u} f = 0$, then $|\EX_{Q+u}f - \EX_{U_n} f|    \geq \EX_{U_n} f$. 
    Therefore,  by Markov Inequality,    $$p \leq \frac{1}{\EX_{U_n} f} \EX_{u\sim U_n} | \EX_{Q+u}f      - \EX_{U_n} f |    .$$  
    Note that $f$ is a symmetric function in the sense that its value on $x$ depends only on the weight $|x|$ of $x$.
    Thus, for any $u\in \{0,1\}^n$,
    $\EX_{Q+u}f = \EX_{\overline{\mu_{Q+u}}}\bar{f}$ and  $\EX_{U_n} f = \EX_{B_n} \bar{f}    $, where $\bar{f}: [0:n] \rightarrow \{0,1\}$ is $1$ iff $w \leq R$ and zero  otherwise.
    Therefore, 
    $$|\EX_{Q+u}f      - \EX_{U_n} f | = |\EX_{\overline{\mu_{Q+u}}}\bar{f}      - \EX_{B_n} \bar{f} | 
\leq 
\| \overline{\mu_{Q+u}} - B_n \|_1.$$
    Noting that $\EX_{U_n} f = \v_n( R)$, we get  
    \[
      p\leq \frac{1}{\v_n(R)} \EX_{u\sim U_n}\| \overline{\mu_{Q+u}} - B_n \|_1. 
\]
    The lemma then  follows from Theorem \ref{wdl1}.
\finito

         \paragraph{Proof of Theorem \ref{brute}} Let $m = n+d$ and 
         extend $C$ to the $\F_2$-linear code $Q = C \times \{0,1\}^d \subset \{0,1\}^{m}$. Thus $Q^\bot = C^\bot \times \vec{0}_J$, where
         $J = \{n+1,\ldots, n+d\}$         and $\vec{0}_J\in \{0,1\}^J$ is the all-zeros vector. By construction, the bilateral minimum distance of $Q^\bot$ is at least $d$ since $d \leq |z| \leq n = m-d$ for each nonzero $z\in Q^\bot$.
         Applying  Corollary \ref{corwcdr}  to $Q$, we get
\begin{equation}\label{qatar2}
  U_m(\B_m(Q; R)) \geq
  1 - \frac{d}{ \v_{n+d}(R)     } \left(e\log{\frac{n+d}{d-1}}\right)^{\frac{d-1}{2}}\left(\frac{d-1}{n+d}\right)^{ \frac{d-5}{4}}. 
\end{equation}
On the other hand, 
\begin{equation}\label{qatar}
  \B_m(Q; R)|_{I} \subset \B_n(C; R), 
         \end{equation}
where $I = \{1,\ldots,n\}$ 
and
$\B_m(Q; R)|_{I}$ is the restriction of
$\B_m(Q; R)$ to $I$. To see why (\ref{qatar}) holds, note that for any
$x\in \B_m(Q; R)$, we have $|x+(y',y'')|\leq R$ for some $y'\in C$ and $y''\in \{0,1\}^d$.
Thus $|x|_I + y'| \leq R$.

The claim then follows from (\ref{qatar}) and (\ref{qatar2}) via  the bounds:
\[
U_n(\B_n(C; R)) \geq
U_n(\B_m(Q; R)|_I) =     U_m(\B_m(Q; R)|_I\times \{0,1\}^J)    \geq  U_m(\B_m(Q; R)). 
\]
\finito

\section{Proofs of Corollary \ref{maincor1} and \ref{maincor2}}\label{corprf}
  The statement of  Corollary \ref{maincor1}  is   restated below for convenience.
  Corollary \ref{maincor2ex} below is a nonasymptotic version of Corollary \ref{maincor2}. 
~\medskip \smallskip \\
              {\bf Corollary \ref{maincor1}}~{\bf (Explicit asymptotic version)}~{\em
Let $C\subsetneq \F_2^n$  be an $\F_2$-linear code whose dual $C^\bot$ has minimum distance at least $d$, where $d\geq 7$ is an odd integer such that $d =o(n)$. 
Let $R = \frac{n}{2}-\Delta$, where 
$$\Delta  = \sqrt{ \frac{1}{13}(d-5)n\log{\frac{n}{d-1}}}.$$
Then, for sufficiently large $n$,  the fraction of points in Hamming cube not covered by $\B_n(C; R)$              is at most
$\left(\frac{d-1}{n}\right)^{\frac{d-5}{13}}.$
                 }\medskip\smallskip\\ \noindent
               {\bf Proof:} Write 
               $d =2t+5$, where $t\geq 1$ is an integer. 
               Let $m=n+d$,  $R' = \frac{m}{2}-\Delta'$, 
               where  $$\Delta' = \sqrt{ \frac{(d-5) m}{12}\log{\frac{m}{d-1}}}-\sqrt{2m}-2,$$
               and  
                 \[
                 p' = \frac{d}{ \v_m(R')     }
 \left(e\log{\frac{m}{d-1}}\right)^{t+2}\left(\frac{d-1}{m}\right)^{ \frac{t}{2}}. 
  \]
               By Theorem \ref{brute}, it is enough to show that 
               that for $n$ large enough,
\begin{equation}  \label{sjfvfd}
  R'\leq R
\end{equation}
  and
  \begin{equation}\label{eessc}
    p' \leq \left(\frac{d-1}{n}\right)^{t/6.5}. 
  \end{equation}   
  Note that since $d = o(n)$, we have 
  $m  = n(1 + o(1))$ and  $d = o(m)$. 

  {\em Proof of (\ref{eessc}):} 
           We have $\Delta' = o(m)$  
           since $ d = o(m)$.  
           To see why
           this holds, note that
$\Delta' \leq
\sqrt{ \frac{(d-1) m}{12}\log{\frac{m}{d-1}}}
= \frac{m}{\sqrt{12}} 
\sqrt{ \frac{d-1}{m}\log{\frac{m}{d-1}}}$ and that  the
function $x \log{\frac{1}{x}}$ is zero at $x=0$.
Hence,   by  Corollary \ref{demltc}, for any $\e>0$ and 
for $m$ large enough 
  \[
           \v_m( R') \geq  e^{-(2+\e)\frac{(\Delta'+\sqrt{2m}+2)^2}{m}}  =      \left(\frac{d-1}{m}\right)^{\frac{(2+\e)t}{6}}. 
           \]
    Therefore
    \[
    p' \leq d \left(e\log{\frac{m}{d-1}  }\right)^{t+2}
    \left(\frac{d-1}{m}    \right)^{ \frac{(1-\e)t}{6}}.  
    \]
    Let $a = \frac{m}{d-1}$ and note that     $a = w(1)$ since $d = o(m)$. 
    We have $d = 2t+5 =2(t+2)+1\leq 2^{t+2}$,
    hence $d \left(e\log{\frac{m}{d-1}  }\right)^{t+2}\leq
    (2e \log{a})^{t+2}    \leq (2e \log{a})^{3t}$. 
    It follows that 
\[
p'
\leq
\left(\frac{\left(2e\log{a}\right)^{\frac{18}{1-\e}}}{a}\right)^{\frac{(1-\e)t}{6}} 
\leq
\left(\frac{1}{a}\right)^{\frac{t}{6.5}}
=
\left(\frac{d-1}{n+d}\right)^{\frac{t}{6.5}}
\leq
\left(\frac{d-1}{n}\right)^{\frac{t}{6.5}},
\]
where the second inequality holds for $\e$ sufficiently small  and 
for  $a$  sufficiently large, i.e.,  for  $n$ sufficiently  large.  

    {\em Proof of (\ref{sjfvfd}):}
    We have
    \[
      R' 
      =\frac{n}{2}-   \sqrt{ \frac{(d-5)(n+d)}{12}\log{\frac{n+d}{d-1}}}      +\sqrt{2(n+d)}+\frac{d}{2}+2. 
\]   
      Since $d = o(n)$, we have
      $\frac{d}{2} +2 = o\left(\sqrt{{(d-5)(n+d)}}\right)$ and 
      $\sqrt{2(n+d)} = o\left(\sqrt{{(n+d)}\log{\frac{n+d}{d-5}}}\right)$,
      hence,  for $n$ large enough,  
\[
R' \leq  \frac{n}{2} - \sqrt{ \frac{(d-5)(n+d)}{13}\log{\frac{n+d}{d-1}}}
\leq
\frac{n}{2} - \sqrt{ \frac{(d-5)n  }{13}\log{\frac{n}{d-1}}}. 
\]
\finito

\begin{corollary}\label{maincor2ex} {\bf (Adding a small code)} 
              Let $C\subsetneq \F_2^n$  be an $\F_2$-linear code whose dual $C^\bot$ has minimum distance at least $d$, where $d\geq 7$ is an odd integer such that $d   = o(n)$.
Then there exists an $\F_2$-linear code $D$ of dimension at most $\ceil{\log_2{n}}$ such that, 
for sufficiently large $n$, the covering radius of $C+D$ is at most
$R = \frac{n}{2}- \sqrt{ \frac{1}{13}(d-5)n\log{\frac{n}{d-1}}}.$
\end{corollary} 
                 {\bf Proof:}  Cohen's argument 
                 is based on iteratively augmenting the code by
                 adding points in $\F_2^n$ to minimize the number of uncovered points (\cite{coh83};  see also  
                 \cite[Section~12.3]{CC97}).
                   Consider the set of points not $R$-covered by $C$, i.e., 
                   $\B_n(C; R)^c$.  Choose $x^{(1)} \in \F_2^n$ to minimize the number of 
                   points not $R$-covered by $C^{(1)} = C \cup (C+x^{(1)})$. 
                   Thus $U_n(\B_n(C^{(1)};R)^c) =U_n(\B_n(C;R)^c \cap (\B_n(C;R)+x^{(1)})^c)$.  By 
                 \cite[Lemma~12.3.1]{CC97}, for each $A \subset \F_2^n$, there exists $x\in \F_2^n$ such that $U_n(A \cap (A+x)) \leq U_n(A)^2$.
                 Thus $U_n(\B_n(C^{(1)};R)^c)                   \leq U_n(\B_n(C;R)^c)^2$.
                 
                   By repeating this process $i$ steps, we get that
                   there exists an $\F_2$-linear code $D$ of dimension $i$ such that $U_n(\B_n(C+D;R)^c)
                   \leq  U_n(\B_n(C;R)^c)^{2^i} < 2^{-2^i}$ 
                   assuming that $n$ is large enough so that $\left(\frac{d-1}{n}\right)^{\frac{d-5}{13}} 
                   < \frac{1}{2}$. Thus,   for $i = \ceil{\log_2{n}}$,
                   $U_n(\B_n(C+D;R)^c)) < 2^{-n}$, i.e., $\B_n(C+D;R)= \{0,1\}^n$. 
                   \finito

                   \section{Extension from codes to distributions} \label{extenap}
  In this section, we simplify a part of the proof in \cite{Baz15}, which makes it possible to 
 extend the results in \cite{Baz15}   and accordingly the results reported in this paper  from codes to distributions.
Namely, we extend the results in \cite{Baz15} on the weight distribution of cosets of codes with bilateral dual distance at least $b$
to translations of bilaterally $k$-wise independent probability distributions, where $k=b-1$. 
In particular, we show that if  $\mu$ is a bilaterally $k$-wise independent  probability distribution on
  $\{0,1\}^n$, then   the average  $L_1$-distance  between the 
  weight distribution of a random translation  of $\mu$ and the binomial distribution  decays quickly in $b$. The decay is exactly the same as in
  \cite{Baz15} with $b-1$   replaced with $k$. This immediately extends the results reported in this paper on the $\varepsilon$-covering radius from
  codes with dual distance $d$ to $k$-wise independent distributions on $\{0,1\}^n$, where  $k=d-1$ and the 
$\varepsilon$-covering radius 
  of a  distribution     is interpreted  as that  of   its support.

 In 
  this section, we  use the weight distributions, Fourier transform, and limited independence 
  notations and definitions given in  Sections \ref{termino}, \ref{frprelS}, \ref{limindep},   respectively.
  We  also need   the following notations for translation and convolution. If $\mu$ is a probability distribution on $\{0,1\}^n$ and $u\in \{0,1\}^n$, 
define $\sigma_u \mu$   to be  the {\em translation}  over $\F_2$ of $\mu$ by $u$, i.e., $(\sigma_u \mu)(x) = \mu(x+u)$.
If $f, g: \{0,1\}^n \rightarrow {\mathbb C}$,  define their {\em convolution} $f\star g: \{0,1\}^n \rightarrow {\mathbb C}$
with respect to addition in $\Z_2^n$  by $(f\star g)(x) =\sum_y f(y)g(x+y)$. 
If $\mu_1,\mu_2$ are probability distributions on $\{0,1\}^n$, their convolution $\mu_1\star \mu_2$ is
a probability distributions on $\{0,1\}^n$;  to sample from  $\mu_1\star \mu_2$,  sample $a\sim \mu_1$, $b\sim \mu_2$,  and return $a+b$.

 In  the proofs of the main results  in \cite{Baz15}, the only part    which relies on  the linearity of the code is the following lemma. 
\begin{lemma}{\bf \cite[Lemma~14]{Baz15}} \label{mainl1}  
If $0\leq \theta < 2\pi$, define $e_\theta:\{0,1\}^n \rightarrow {\mathbb C}$ by $e_\theta(x) = e^{i\theta |x|}$. 
Let $Q\subsetneq \F_2^n$  be an $\F_2$-linear code and $0\leq \theta < 2\pi$. 
Then 
\[
\EX_{u\sim U_n} |\EX_{\mu_{Q+u}} e_\theta - \EX_{U_n} e_{\theta}|^2 = \EX_{y\sim \mu_Q} (\cos{\theta})^{|y|}- \left(\frac{\cos{\theta}+1}{2}\right)^n.
\]
\end{lemma}
Lemma \ref{mainl1} is used in the proof of \cite[Theorem~5]{Baz15}: 
\begin{theorem} {\bf    \cite[Theorem~5]{Baz15}}\label{th5b15} {\bf (Mean-square-error bound)} 
Let $Q\subsetneq \F_2^n$  be an $\F_2$-linear code whose dual $Q^\bot$ has bilateral minimum distance at least $b=2t+1$, where $t\geq 1$ is an integer. 
Then,
for each $0\leq \theta < 2\pi$, we have the  bounds: 
\begin{itemize}
\item[a)]  {\em $($Small dual distance bound$)$} 
\[
   \EX_{u\sim U_n} | \EX_{\mu_{Q+u}} e_\theta - \EX_{U_n} e_{\theta}|^2 
\leq \left(e\log{\frac{n}{2t}}\right)^{2t}\left(\frac{2t}{n}\right)^{t} 
\]
\item[b)]  {\em $($Large dual distance bound$)$} 
\[
   \EX_{u\sim U_n} | \EX_{\mu_{Q+u}} e_\theta - \EX_{U_n} e_{\theta}|^2 
  \leq 2 e^{-\frac{t}{5}}.
\]
\end{itemize}
\end{theorem}
Lemma \ref{mainl1e} below extends Lemma \ref{mainl1} from codes to probability distributions and it has a simpler proof. 
\begin{lemma}\label{mainl1e} 
Let $\mu$ be a  probability distribution on $\{0,1\}^n$ 
and $0\leq \theta < 2\pi$. 
Then 
\[
\EX_{u\sim U_n} |\EX_{\sigma_u \mu} e_\theta - \EX_{U_n} e_{\theta}|^2
   = \EX_{y\sim \mu\star \mu} (\cos{\theta})^{|y|} -\EX_{y\sim U_n} (\cos{\theta})^{|y|}. 
   \]
\end{lemma}
   Note that $\EX_{y\sim U_n} (\cos{\theta})^{|y|} = \left(\frac{\cos{\theta}+1}{2}\right)^n$.  Moreover, if $\mu = \mu_Q$, where $Q\subset \F_2^n$ is  an $\F_2$-linear code, then $\mu_Q\star \mu_Q  = \mu_Q$.
  \paragraph{Proof}  
  We have 
    \[
    \EX_{u\sim U_n} |\EX_{\sigma_u \mu} e_\theta - \EX_{U_n} e_{\theta}|^2
    =  \EX_{u\sim U_n} |\EX_{\sigma_u \mu} e_\theta|^2 - |\EX_{U_n} e_{\theta}|^2 
    \]
    and
    \begin{eqnarray*}
      \EX_{u\sim U_n} |\EX_{\sigma_u \mu} e_\theta|^2 &=&\EX_{u\sim U_n} |\sum_x  \mu(x) e_\theta(x+u)|^2 \\
      &=&
      \EX_{u \in \{0,1\}^n} \sum_{x,y} \mu(x)\mu(y) e_\theta(u+x)\overline{e_\theta(u+y)}\\
      &=&       \sum_{x,y} \mu(x)\mu(y)       \EX_{u \in \{0,1\}^n} e_\theta(u)\overline{e_\theta(u+x+y)}\\
      &=&       \EX_{\mu\star \mu}    e_\theta \ostar \overline{e_\theta}, 
    \end{eqnarray*}
    where $\ostar$ is the {\em weighted convolution} operator:
    if $f, g: \{0,1\}^n \rightarrow {\mathbb C}$,  their weighted convolution
$f\ostar g = \frac{1}{2^n} f \star g$, i.e., $(f\ostar g)(x) = \EX_y f(y)g(x+y)$, hence $\widehat{f \ostar g} = \widehat{f} \widehat{g}$.

    Then the Lemma follows from the fact that 
    \begin{equation}\label{econv}
      (e_\theta \ostar \overline{e_\theta})      (x)    = (\cos{\theta})^{|x|}.  
    \end{equation}
    Note that  $\EX_{U_n}e_\theta \ostar \overline{e_\theta} = |\EX_{U_n} e_{\theta}|^2    $. Thus,
    by (\ref{econv})    , $|\EX_{U_n} e_{\theta}|^2 = \EX_{y\sim U_n} (\cos{\theta})^{|y|}$. 

    To verify (\ref{econv}), we go to the Fourier domain. In the Fourier domain, (\ref{econv}) is equivalent to  
    $\widehat{e_\theta \ostar \overline{e_\theta}}    = \widehat{g_{\cos{\theta}}}$, where 
    $g_r(x) = r^{|x|}$. Since $\widehat{f\ostar g} = \widehat{f}\widehat{g}$, we have to show that
    $|\widehat{e_\theta}|^2    = \widehat{g_{\cos{\theta}}}$.
    We need the following basic lemma about the Fourier transform of exponential function, e.g.,
    \cite[Lemma~11]{Baz15}:
\begin{lemma}\label{frprdcmp}
 Let $r$ be complex number and  
$g_r:\{0,1\}^n\rightarrow \CN$ be given by $g_r(x) = r^{|x|}$.
Then 
$
     \widehat{g_r}(z) =\left( \frac{1+r}{2}\right)^n \left(\frac{1-r}{1+r}\right)^{|z|}.
$ 
Moreover, if $r= e^{i\theta}$, then $\widehat{g_{r}}(z) =e^{i n\theta/2} \left(\cos\frac{\theta}{2}\right)^{n} \left(-i \tan{\frac{\theta}{2}}\right)^{|z|}$.
\end{lemma}
Therefore 
\[
  |\widehat{e_\theta}(z)|^2=
  \left(\cos\frac{\theta}{2}\right)^{2n} \left(\tan{\frac{\theta}{2}}\right)^{2|z|}
  =\left( \frac{1+\cos{\theta}  }{2}\right)^n \left(\frac{1-\cos{\theta}}{1+\cos{\theta}}\right)^{|z|} = g_{\cos{\theta}}(z), 
  \]
  where the second equality follows from  the trigonometric identities
  $\left(\cos\frac{\theta}{2}\right)^{2}
  =\frac{1+\cos{\theta}  }{2}$ and 
  $\left(\tan{\frac{\theta}{2}}\right)^{2}
  =\frac{1-\cos{\theta}}{1+\cos{\theta}}$.  
\finito

As explicitly noted in \cite[Section V~p.~6499]{Baz15}, the proof of Theorem \ref{th5b15}  bounds the term $\EX_{y\in \mu_Q} (\cos{\theta})^{|y|} -\left(\frac{\cos{\theta}+1}{2}\right)^n$  
by ignoring the linearity of $Q$ and  using only the bilateral $k$-wise independence property of $\mu_Q$, where $k=b-1$.
This property holds also for $\mu \star \mu$; 
if $\mu$ is bilaterally $k$-wise independent,  then  so is $\mu\star \mu$.  
This follows from the definition of bilateral $k$-wise independence  since  $\EX_{\mu \star  \mu } \X_z  =\left(\EX_{\mu } \X_z \right)^2$, for each $z\in \{0,1\}^n$. 

Accordingly, using Lemma \ref{mainl1e},
Theorem \ref{th5b15} as well as  
\cite[Theorem ~2]{Baz15} ($L_\infty$-bound)
and \cite[Corollary~3]{Baz15} ($L_1$-bound, i.e.,  Theorem \ref{wdl1} in this paper)  
extend as follows from codes with bilateral minimum distance at least $b$ to
bilaterally $k$-wise independent probability distributions, where $k=b-1$.

 \begin{theorem}   \label{maino}
   {\bf (Bilateral limited independence versus weight distribution of translates; mean-square-error bound)}
 Let $\mu$ be a  bilaterally $k$-wise independent probability distribution on $\{0,1\}^n$, where $k \geq 2$ is an even integer.  
 Then,
for each $0\leq \theta < 2\pi$, we have the  bounds: 
\begin{itemize}
\item[a)]  {\em $($Small $k$ bound$)$} 
\[
   \EX_{u\sim U_n} | \EX_{\sigma_u \mu} e_\theta - \EX_{U_n} e_{\theta}|^2 
\leq \left(e\log{\frac{n}{k}}\right)^{k}\left(\frac{k}{n}\right)^{\frac{k}{2}}.  
\]
\item[b)]  {\em $($Large $k$ bound$)$} 
\[
   \EX_{u\sim U_n} | \EX_{\sigma_u \mu} e_\theta - \EX_{U_n} e_{\theta}|^2 
  \leq 2 e^{-\frac{k}{10}}.
\]
\end{itemize}
 \end{theorem}
 \begin{theorem}\label{sumth}{\bf (Bilateral limited independence versus weight distribution of translates; $L_\infty$-bound)}
 Let $\mu$ be a  bilaterally $k$-wise independent probability distribution on $\{0,1\}^n$, where $k \geq 2$ is an even integer. 
 Then, we have the bounds:   
\begin{itemize}
\item[a)] {\em $($Small $k$ bound$)$} 
\[
   \EX_{u\sim U_n} \| \overline{\sigma_u \mu} - B_n\|_\infty
\leq \left(e\log{\frac{n}{k}}\right)^{\frac{k}{2}}\left(\frac{k}{n}\right)^{\frac{k}{4} }. 
\]

\item[b)] {\em $($Large $k$ bound$)$} 
\[
   \EX_{u\sim U_n} \| \overline{\sigma_u \mu} - B_n\|_\infty  \leq \sqrt{2} e^{-\frac{k}{20}}.
\]
\end{itemize}
 \end{theorem}
 \begin{theorem} \label{wdle1} {\bf (Bilateral limited independence versus weight distribution of translates; $L_1$-bound)}  
 Let $\mu$ be a  bilaterally $k$-wise independent probability distribution on $\{0,1\}^n$, where $k \geq 6$ is an even integer. 
 Then, we have the bounds:  
\begin{itemize}
\item[a)]  {\em $($Small $k$ bound$)$} 
  $$\EX_{u\sim U_n} \| \overline{\sigma_u \mu}  - B_n \|_1  \leq (k+1)\left(e\log{\frac{n}{k}}\right)^{\frac{k}{2}}\left(\frac{k}{n}\right)^{ \frac{k}{4}-1}.$$
\item[b)] {\em $($Large $k$ bound$)$} 
   $$\EX_{u\sim U_n} \| \overline{\sigma_u \mu} - B_n \|_1  \leq \sqrt{2}(n+1)e^{-\frac{k}{20}}.$$ 
\end{itemize}
\end{theorem}
As in \cite{Baz15},  Theorem \ref{wdle1} follows from  Theorem \ref{sumth}, which in turns follows from  Theorem \ref{maino}.

Accordingly, Theorem \ref{brute}  (Nonasymptotic bound) and Corollaries \ref{maincor1} (Explicit asymptotic version) and \ref{maincor2ex} (Adding a small code)    extend as follows from codes with minimum distance at least $d$ to
$k$-wise independent probability distributions, where $k=d-1$.
\begin{definition} {\bf ($\varepsilon$-covering radius of a probability distribution)} 
      Let $\mu$ be a probability distribution
      on $\{0,1\}^n$.
      The {\em covering radius} of  $\mu$ is the covering radius of its support.
  Equivalently,  the covering radius of $\mu$ is the minimum $r$
  such that $\mu(\B_n(x; r))\neq 0$ for each $x\in \{0,1\}^n$.

  More generally, if  $0\leq \varepsilon \leq 1$,  
  the {\em $\varepsilon$-covering radius} of  $\mu$ 
  is the  {\em $\varepsilon$-covering radius} of  its support. Equivalently,  the $\varepsilon$-covering radius of $\mu$ is the minimum $r$ such that the fraction of points
  $x\in \{0,1\}^n$   such that $\mu(\B_n(x; r))= 0$ is at most $\varepsilon$. 
\end{definition}
  \begin{theorem}\label{brutee}
    {\bf (Limited independence versus almost-all-covering radius)} 
 Let $\mu$ be a $k$-wise independent probability distribution on $\{0,1\}^n$, where $k \geq 6$ is an even integer. 
  If $R>0$, let  
\[
\varepsilon  = \frac{d}{ \v_{n+k+1}(R)     } \left(e\log{\frac{n+k+1}{k}}\right)^{\frac{k}{2}}\left(\frac{k}{n+k+1}\right)^{ \frac{k}{4}-1}  
\]
and assume that  $\varepsilon \leq 1$. Then the  $\varepsilon$-covering radius of $\mu$ is most $R$.
  \end{theorem}
To adapt  the proof of Theorem \ref{brute} into a the setup  of Theorem \ref{brutee}, given
 a $k$-wise independent probability distribution $\mu$ on $\{0,1\}^n$,   consider the probability distribution 
 $\gamma = \mu \times U_d$ on $\{0,1\}^m$, where $d=k+1$ and $m=n+d$.  Then  $\gamma$  is bilaterally $k$-wise independent. 
The reason is that  if $z\in \{0,1\}^m$ is such that $|z|>  m-d=n$,
then with $I = \{1,\ldots,n\}$ and  $J = \{n+1,\ldots, n+d\}$,  we have 
$z|_{J} \neq 0$, hence $\EX_{U_d} \X_{z|_J}= 0$, and accordingly $\EX_\gamma \X_z = 0$ since 
$\EX_\gamma \X_z = \EX_\mu \X_{z|_I} \EX_{U_d} \X_{z|_J}$.

  \begin{corollary}\label{maincor1e}
    {\bf (Explicit asymptotic version)}
    Let $\mu$ be a $k$-wise independent probability distribution on $\{0,1\}^n$, where $k \geq 6$ is an even integer such that $k =o(n)$. 
              Then, for sufficiently large $n$, the $\left(\frac{k}{n}\right)^{\frac{k-4}{13}}$-covering radius of $\mu$ is at most $R = \frac{n}{2}-\Delta$, where
              $$\Delta  = \sqrt{ \frac{1}{13}(k-4)n\log{\frac{n}{k}}}.$$
  \end{corollary}
  \begin{corollary}\label{maincor2e} {\bf (Convolution with a small code)}
         Let $\mu$ be a  $k$-wise independent probability distribution on $\{0,1\}^n$, where $k \geq 6$ is an even integer such that $k =o(n)$. 
Then there exists an $\F_2$-linear code $D$ of dimension at most $\ceil{\log_2{n}}$ such that, 
for sufficiently large $n$, the covering radius of  $\mu\star \mu_D$ is at most                $R = \frac{n}{2}- \sqrt{ \frac{1}{13}(k-4)n\log{\frac{n}{k}}}.$
  \end{corollary}

\section{Open problems}\label{openprobs}
We conclude with the following  open questions: 
\begin{itemize}
\item 
  As noted in the introduction, after Corollary \ref{maincor1}, the upper bound  
  of
  Corollary \ref{maincor1}
  on $R-\frac{n}{2}$,   
  where $R$ is the almost-all covering radius,
                        is asymptotically tight up to a factor of $\sqrt{\frac{13}{2}}$ 
                        in comparison
                        to random linear codes (see Lemma \ref{GVCSCA}). 
  The proofs of Theorem \ref{brute} and Corollary \ref{maincor1} can be easily tuned to bring 
  the $\sqrt{\frac{13}{2}}$ factor down to  $2+\e$, for any $\e>0$.
 The gain is at the cost of   increasing the fraction of uncovered points while keeping it $o(1)$. 
 Is it possible to go below $2$?
           \item Corollary \ref{maincor1}  assumes that the dual distance $d$ is at least $7$.
             Is it possible to extend it to smaller values of $d$?

             \item   
                 Consider the  block-length-$n$ dual BCH code $C=BCH(s,m)^\bot$, where 
                 $m\geq 2$ is an integer, $n = 2^m -1$, 
                 and $s$ is an  integer such that $2s-2 < 2^{m/2}$.
                 If $s\geq 3$, we know from     Corollary  
                 \ref{maincor2ex} 
that there exists a small $\F_2$-linear code $D$ of dimension at most $\ceil{\log_2{n}}=m$ such that, 
for sufficiently large $n$, the covering radius of $C+D$ is at most  $\frac{n}{2}- \sqrt{ \frac{1}{13}(2s-4)n\log{\frac{n}{2s}}} =
\frac{n}{2}- \Theta(\sqrt{sn\log{n}})$.
It would be interesting to   explicitly construct 
such a code $D$ using  algebraic tools. 
\end{itemize}

\appendix

\paragraph{\Large Appendix}

\section{Proof of Corollary \ref{cohcor}}\label{appb}
The corollary  is restated below for convenience. 
\medskip \\
         {\bf Corollary \ref{cohcor}}~{\em
           For each $\e>0$, there exists $\delta>0$ such that the following holds. 
    Let $n\geq 1$ be an integer and $s> 1$ be  such that $s\log_2{n} \leq \delta n$. 
    Then for $n$ large enough, 
    there exists an $\F_2$-linear code $C\subset \F_2^n$ of dimension at most     $\ceil{s \log_2{n}}$ and 
    covering radius 
\[
R \leq \frac{n}{2} - \sqrt{\frac{ (s-1) n\log{n} }  {2+\e}} + \sqrt{2n}+2. 
\]}
\noindent Let $\Delta = \sqrt{\frac{ (s-1) n\log{n} }  {2+\e}} - \sqrt{2n}-2$. By Theorem \ref{cohthm}, it is enough to show that 
$\log_2{  \frac{n (\log{2})}{\v_n( \frac{n}{2}    -\Delta)}} \leq     s \log_2{n}$, i.e.,
$\v_n( \frac{n}{2}-\Delta) \geq  \frac{\log{2}}{n^{s-1}}$.  Since
$s \log_2{n} \leq \delta n$, we have
$\Delta \leq   \sqrt{\frac{\delta\log{2}}{2+\e}} n$. Applying 
Lemma  \ref{demltc}, we get that 
for sufficiently small  $\delta$ and sufficiently large $n$, 
\[
           \v_n( {n/2}-\Delta) \geq  e^{-(2+\e)\frac{(\Delta+\sqrt{2n}+2)^2}{n}} = \frac{1}{n^{s-1}} >  \frac{\log{2}}{n^{s-1}}.
 \]

\section{Proof of Lemma \ref{GVCSCA}}\label{appa}       
The lemma is restated below for convenience.
\medskip \\
         {\bf Lemma \ref{GVCSCA}}~{\em
        Consider any  $0\leq \varepsilon < 1$ and let             $n\geq 1$  be an integer   and            $ n \leq K \leq 2^{n-1}$  be  an integer power of $2$.   
           Then, all but at most   $\frac{1}{n}$ fraction of 
           $\F_2$-linear codes $C\subset \F_2^n$ of size $K$ have  $\varepsilon$-covering radius 
\[
R \geq  
\frac{n}{2} - \sqrt{
  \frac{dn}{2}  \log{\frac{en}{d}} +   n \log{\frac{n+1}{1-\varepsilon}}
},        \]
where $d$ is the minimum distance  of $C^\bot$.
         }         
  \smallskip \noindent

\noindent 
  We need the following simple variation of the sphere-covering bound: 
  \begin{lemma}\label{scv} {\bf (Sphere-covering bound adaptation to almost-all covers)} 
Let $0\leq \varepsilon < 1$ and     $n\geq 1$. 
    Then for any code $C \subset \{0,1\}^n$ of size $K$, where $K \geq 1$, 
    the $\varepsilon$-covering radius of $C$ is at least
    \[ 
R \geq  \frac{n}{2} - \sqrt{\frac{1}{2} n \log{\frac{K}{1-\varepsilon}} }.   
\]
\end{lemma}
  The proof of Lemma \ref{scv} follows from exactly the  same counting argument used  to
  establish the sphere-covering bound 
  \cite[Theorem~12.5.1]{CC97}.

The upper bound on $K$ in terms of $d$ comes from  
Gilbert-Varshamov bound.
Choose the generator matrix $G_{k^\bot\times n}$ of the dual code $C^\bot$ uniformly at random, where $k^{\bot} = n - \log_2{K}$. Let $d$ be the minimum distance of
$C^\bot$ and let $1\leq  d_0\leq \frac{n}{2}+1$ be an integer. 
The probability that $d < d_0$  is 
at most
$$(|C^\bot|-1)2^{-n}|\B_n(0; d_0-1)|  \leq \frac{d_0}{K} \binom{n}{d_0-1}  \leq
\frac{d_0}{K}
\left(\frac{en}{d_0-1}\right)^{d_0-1}
\leq \frac{1}{n}$$ 
if $K \geq f(d_0)$, where 
$f(x) =   n x\left(\frac{en}{x-1}\right)^{x-1}$.    
Since $f(x)$ is increasing in $x$
for all $1 \leq x \leq n+1$, we conclude that, with probability at least $1-\frac{1}{n}$, 
$d \geq \floor{f^{-1}(K)}$ if $1\leq  \floor{f^{-1}(K)} \leq \frac{n}{2}+1$, 
hence $K \leq f(d+1)$ if $f(1) \leq K \leq f(\frac{n}{2}+1)$.
We have $f(1) = n \leq K$  and $f(\frac{n}{2}+1) >2^{n}$ for all $n\geq 1$.
Therefore, 
\[ 
R \geq  \frac{n}{2} - \sqrt{\frac{1}{2} n \log{
    \left(\frac{n (d+1)}{1-\varepsilon}    \left(\frac{en}{d}\right)^{d} \right)}} \geq 
\frac{n}{2} - \sqrt{
  \frac{dn}{2}  \log{\frac{en}{d}} +   n \log{\frac{n+1}{1-\varepsilon}}.
  }
\]

\section*{Acknowledgment}
The author would like to thank the anonymous reviewers for their useful and constructive comments.

\end{document}